\documentclass[dvips]{article}
\usepackage{icrctc07}

\title{Detection of Cherenkov light from air showers with Geiger-APDs}
\shorttitle{Detection of air showers with Geiger-APDs}
\authors{A. N. Otte$^{1,2}$, I. Britvitch$^{3}$, A. Biland$^{3}$,  F. Goebel$^{1}$, E. Lorenz$^{3}$, F. Pauss$^{3}$, D. Renker$^{4}$, U. R\"oser$^{3}$, T. Schweizer$^{1}$}
\shortauthors{A. N. Otte and et al}
\afiliations{ $^1$Max-Planck-Institut f\"ur Physik, D-80805
M\"unchen, Germany\\ $^2$ Humboldt-Universit\"at zu Berlin,
D-12489 Berlin, Germany\\
 $^3$
ETH Zurich, CH-8093 Zurich, Switzerland\\
$^4$ Paul Scherrer Institut, CH-5232 Villigen, Switzerland}
\email{otte@mppmu.mpg.de}

\abstract{We have detected Cherenkov light from air showers with
Geiger-mode APDs (G-APDs). G-APDs are novel semiconductor
photon-detectors, which offer several advantages compared to
conventional photomultiplier tubes in the field of ground-based
$\gamma$-ray astronomy.  In a field test with the MAGIC telescope
we have tested  the efficiency of a G-APD / light catcher setup to
detect Cherenkov light from air showers. We estimate a detection
efficiency, which is 60\% higher than the efficiency of a MAGIC
camera pixel. Ambient temperature dark count rates of the tested
G-APDs are below the rates of the night sky light background.
According to these recent tests G-APDs promise a major progress in
ground-based gamma-ray astronomy.}

\begin{document}
\maketitle

\section{Introduction and Motivation}

Ground-based $\gamma$-ray astronomy is a rapidly expanding and
successful field in astroparticle physics. The Very High Energy
(VHE) $\gamma$-ray astronomy window was opened by the Whipple
collaboration with the detection of the Crab Nebula at energies
above 1 TeV using the imaging air Cherenkov telescope (IACT)
technique \cite{1989ApJ...342..379W}. An IACT records by means of
a finely segmented PMT camera the very weak Cherenkov light
flashes from VHE $\gamma$-ray induced air showers. Since 1989 more
than 70 sources of VHE-$\gamma$-rays have been detected. Most of
them have been discovered in the last four years by the most
recent generation of IACTs.

 The MAGIC telescope is an IACT with a
17\,m diameter reflector. It is currently the largest and
technologically most advanced IACT. MAGIC is aiming to uncover the
so far nearly unexplored VHE $\gamma$-ray domain between 30 and
150\,GeV, important for the study of several fundamental physics
questions such as the extragalactic background light, gamma ray
bursts, pulsars, dark matter and tests of quantum gravity.

Increasing the sensitivity and/or lowering the threshold in this
important energy region requires telescopes with a higher
collection efficiency for Cherenkov photons. While it seems
difficult and expensive to increase the reflector area much beyond
the area of existing telescopes, it is more promising to develop
photon detectors with higher photon detection efficiencies (PDE).

Since a few years, a new type of semiconductor photon detector is
being developed with single photon resolution and a potential for
much higher PDEs than that of PMTs. The so-called G-APD is now in
the transition from an R\&D device to a commercial product. The
size of  some G-APDs is now sufficiently large to evaluate their
possible use in IACTs.

We carried out some first tests, which showed that Cherenkov light
from air showers can be detected with latest G-APDS
\cite{britvich}. After these promising tests we have placed a
matrix of four G-APDs in the focal plane of the MAGIC telescope.
The aim of this test was to compare the G-APD to a MAGIC camera
pixel in terms of their efficiency to detect Cherenkov light. The
results of this field test are presented here.

\begin{figure}[htb]

\begin{center}

\includegraphics*[width=0.9\columnwidth]{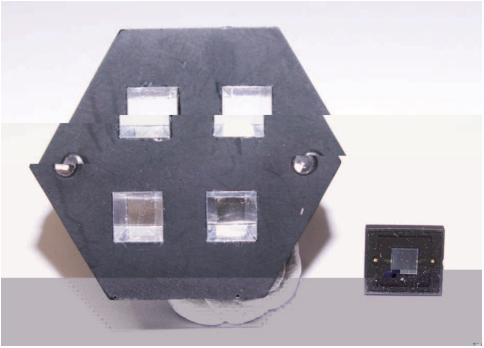}
                    \caption{Front view of the matrix of four light catchers with entrance areas of $6\times6\,$mm$^2$ and exit areas of $3\times3\,$mm$^2$.
                    One of the tested $3\times3\,$mm$^2$ MPPCs is shown in the lower right side. In the test, this MPPC was coupled to the exit area of one of the light catchers.}
               \label{mppc_matrix}
               \end{center}
\end{figure}

\section{The G-APD}

The G-APD or SiPM, MPPC, MRS-APD, ... was initially developed by
several Russian groups
\cite{2003NIMPA.504...48B,1995NIMPA.367..212B,2004NIMPA.518..560G}.
In a G-APD typically 100 to 1000 small avalanche photo diodes
(cells) are integrated per mm$^2$ detector area. Each cell
operates autonomous in limited Geiger mode. The output of a G-APD
is the sum signal of all cells.

The main advantage of G-APDs employed in astroparticle physics is
the potential high PDE, which can be up to a factor of three
higher than that of classical PMTs. Other advantages are:
\begin{itemize}
 \item{single photoelectron response}
 \item{compactness}
 \item{insensitivity to magnetic fields}
 \item{low operation voltage ($<100\,$V) and high gains
 ($10^5-10^6$)}
 \item{no damage when exposed to sunlight, even when under bias}
 \item{long-term prospects for low fabrication costs}
\end{itemize}

Disadvantages of G-APDs are:

\begin{itemize}
  \item{current sensors sizes are limited to
  \mbox{$<10\,$mm$^2$}}
  \item{dark noise rates between 100\,kHz and several MHz per
  mm$^2$ sensor area at room temperature}
  \item{so-called optical crosstalk, i.e.~correlated firing of several cells}
\end{itemize}
%

For our tests we have used four $3\times3$\,mm$^2$ prototype
G-APDs from Hamamatsu; MPPCs with about 50\% peak PDE in the blue
spectral range, cell sizes of $50\times50\,\mu$m$^2$, $\sim2\,$MHz
dark rate at room temperature, and signal rise- and fall-times of
4\,ns and 100\,ns respectively.

\section{Detection of Cherenkov light using the MAGIC telescope}

For the detection of Cherenkov light using the MAGIC telescope we
grouped the aforementioned 4 MPPCs in a 2 $\times$ 2 matrix
(s.~Figure \ref{mppc_matrix}). The effective area of each sensor
was enhanced by a compound miniature light concentrator. The first
stage is a light catcher made of a high reflective foil (95\%
reflectivity for wavelengths down to 320\,nm). The second stage is
a light catcher made of plexiglas.

\begin{figure}[htb]

\begin{center}

\includegraphics*[width=0.85\columnwidth]{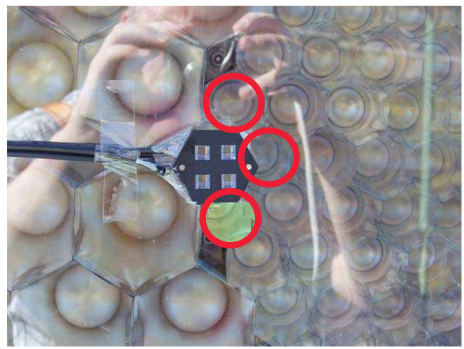}
                    \caption{MPPC matrix mounted onto the entrance window of the MAGIC camera. The three
                     red circles mark the PMTs, whose signals the MPPC signals were compared to.}
               \label{mppc_mounted}
               \end{center}
\end{figure}

\begin{figure*}[t!]
\begin{center}

\includegraphics*[width=0.9\textwidth]{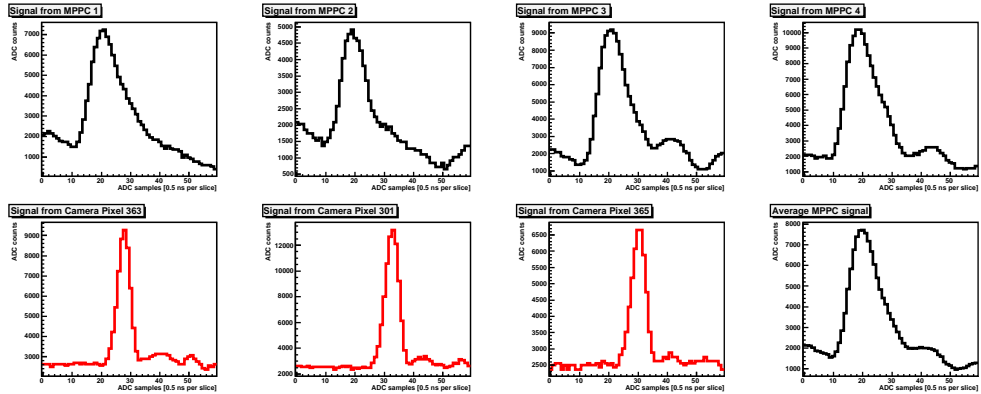}
                    \caption{The upper row shows the signals of the four MPPCs for the event shown in Figure \ref{event}.
                    The average over the four MPPCs is shown in the
                    rightmost
                    figure in the lower row. The three recorded signals shown in the lower row in red
                     are from  the three PMTs surrounding the MPPC array.}
               \label{signals}
               \end{center}
\end{figure*}

The MPPC-matrix was mounted onto the entrance window of the MAGIC
camera lined up with one pixel in the outermost ring of the finer
pixelated inner camera (s.~Figure \ref{mppc_mounted}). The MPPC
signals were amplified with MAR 8-ASM amplifiers from
Mini-Circuits. After the amplifier the signals were clipped with a
2\,ns long cable to avoid the \mbox{$>100\,$ns} long tails of the
MPPC signals, i.e., to minimize baseline shifts and pile-up. The
analog signals were then sent via 160\,m long optical links into
the counting house were they were digitized by the MAGIC-DAQ
\cite{goebeldaq} at a sampling frequency of 2\,GSamples/s each
time the readout was triggered by a cosmic ray shower.

In this configuration each MPPC recorded a single photon rate from
the light of night sky of about 25\,MHz. This rate is about 10
times higher than the intrinsic dark rate of the MPPC at room
temperature. Thus in future IACT applications of G-APDs, only
moderate requirements are needed in terms of temperature
stabilization and cooling.

\begin{figure}[htb]

\begin{center}
\includegraphics*[width=0.8\columnwidth]{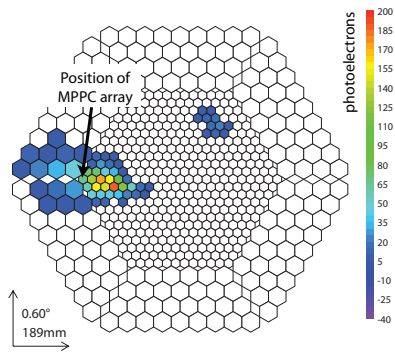}
                    \caption{One event recorded during the test. The corresponding signals of the
                    MPPCs and of the three  PMTs surrounding the MPPC array are shown in Figure \ref{signals}. Mark position of the MPPC array.}
               \label{event}
               \end{center}
\end{figure}

Figure \ref{event} shows one image of an air shower that was
recorded during the test. For this event clear signals are seen by
all four MPPCs (see upper row of Figure \ref{signals}). The
signals recorded by the three PMTs closest to the MPPC-array (red
circles in Figure \ref{mppc_mounted}) are shown by the three red
graphs in the lower row of the Figure \ref{signals}.

In the above event each MPPC recorded on average 12 photoelectrons
and the surrounding PMTs on average 130 photoelectrons. Note that
the entrance area of the light catcher of a PMT and a MPPC differ
by a factor 20. Therefore, the PMT signals have to be scaled down
by this factor in order to allow for a quantitative comparison
between both sensors, i.e.~$130/20\sim6$\,photoelectrons, which is
half the intensity recorded by the MPPCs.

\begin{figure}[htb]

\begin{center}

\includegraphics*[width=\columnwidth]{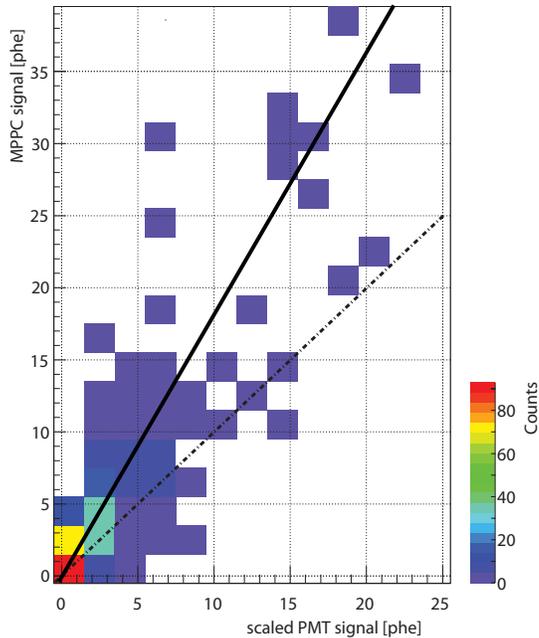}
                    \caption{Comparison of the MPPC and PMT photoelectron signal scaled to the MPPC area.}
               \label{correlation}
               \end{center}
\end{figure}

For about 300 events we calculated the photoelectron intensity
recorded by the MPPCs and the area scaled intensity recorded by
the surrounding PMTs. Figure \ref{correlation} shows the
correlation between the MPPC and PMT recorded intensities. In the
figure the number of photoelectrons recorded by the MPPCs is above
the dashed line through the origin (slope 1) for most of the
events, which indicates that the MPPCs have a  higher PDE than the
PMTs.

On average the MPPCs record a 80\% higher photoelectron intensity
than the PMTs, indicated by the solid line in Figure
\ref{correlation} (slope 1.8). However, this value is not
corrected for optical crosstalk, which results in an overestimate
of the PDE of the MPPCs. After correcting for optical crosstalk we
estimate that the MPPC light catcher arrangement is 60\% more
sensitivity than the PMT light catcher arrangement used in MAGIC.

Note that above we not only estimated the PDE of the MPPC. Instead
the combined efficiency of the photondetector and the light
catcher was estimated. The MPPC light catcher was not optimized,
which leaves room for further improvement.

\section{Discussion and Conclusion}

In summary we can conclude the following results:

\begin{itemize}
 \item{It is possible to detect
  Cherenkov light from air showers with available G-APDs}
  \item{With the tested G-APDs, we obtained a significant increase in the efficiency to collect Cherenkov photons from air showers
  compared to a MAGIC PMT-pixel ($\approx60\%$)}
  \item{The intrinsic noise rate of the tested G-APDs at room temperature was a factor of 10 below the
  level of the night sky background light}
\end{itemize}

As a consequence of the significant increase in PDE we conclude
that by means of G-APDs one can either increase the sensitivity
and lower the threshold of IACTs or one can reduce the reflector
area of telescopes while conserving the same performance in case
classical PMTs are used.

Even if available G-APD already show impressive characteristics
further improvements are desirable like larger sensor areas of
$5\times5\,$mm$^2$ up to \mbox{$10\times10$\,mm$^2$}, a reduction
of optical crosstalk to well below 5\%, and a further reduction of
the intrinsic noise to \mbox{$<100\,$kHz/mm$^2$}.

\bibliography{icrc1070}
\bibliographystyle{unsrt}

\end{document}